\documentstyle[epsfig,amssymb]{mn}
\title[Cold and warm dust along a merging galaxy sequence]  
{Cold and warm dust along a merging galaxy sequence}
 
\author[Xilouris, Georgakakis, Misiriotis \& Charmandaris]
  {E. M. Xilouris$^{1}$, A. E. Georgakakis$^{1}$, A. Misiriotis$^2$, V. Charmandaris$^3$ \\  
  $^1$ Institute of Astronomy \& Astrophysics, National Observatory of
  Athens, I. Metaxa \& V. Pavlou, Athens, 15236, Greece \\ 
  $^2$University of Crete, Physics Department, PO Box 2208, 71003 Heraklion, Crete, Greece\\
  $^3$ Astronomy Department, Cornell University, Ithaca, NY 14853\\}
\begin{document}
\maketitle  
 
\begin{abstract}
We investigate the cold and warm dust properties during galaxy
interactions using a merging galaxy sample ordered into a
chronological sequence from  pre- to post-mergers. Our sample
comprises a total of 29 merging systems selected to  have far-infrared
and sub-millimeter observations. The sub-millimeter data
are mainly culled from the literature while for 5
galaxies (NGC\,3597, NGC\,3690, NGC\,6090, NGC\,6670 and NGC\,7252)
the sub-millimeter observations are presented here for the first
time. We use the 100--to--$\rm 850\mu m$ flux density ratio,
$f_{100}/f_{850}$, as a proxy to the mass fraction of the warm and the
cold dust in these systems. We find evidence for an increase in
$f_{100}/f_{850}$ along the merging sequence from early to advanced
mergers and interpret this trend as an increase of the warm relative
to the  cold dust mass. We argue that the two key parameters affecting
the $f_{100}/f_{850}$ flux ratio is the star-formation rate and the
dust content of individual systems  relative to the stars. Using a
sophisticated model for the absorption and re-emission of the stellar
UV radiation by dust we show that these parameters can indeed explain
both the increase and the observed scatter in the  $f_{100}/f_{850}$
along the merging galaxy sequence. We also discuss  our results under
the hypothesis that elliptical galaxies are formed via  disc galaxy
mergers.
\end{abstract}


%%%%%%%%%%%%%%%%%%%%%%%%%%%%%%%%%%%%%%%%%%%%%%%%%%%%%%%%
\section{Introduction}\label{sec_intro}
%%%%%%%%%%%%%%%%%%%%%%%%%%%%%%%%%%%%%%%%%%%%%%%%%%%%%%%%

Early studies on the dust properties of spiral galaxies using IRAS
data suggested a gas--to--dust ratio about 5--10 times larger than the
Milky Way (Devereux \& Young 1990; Sanders et al. 1991). This apparent
conflict was interpreted as indirect evidence for the presence of a
cold dust component at a temperature T=15-25\,K heated by the diffuse
interstellar radiation field (Cox, Krugel \& Mezger 1986;
Rowan-Robinson \& Crawford 1989). Dust at such low temperatures
remains undetected at the  IRAS wavebands, that primarily probe warmer
dust ($\rm T>30$\,K), and can only be identified by observations at
longer wavelengths ($\rm >100\mu$m).  It was only recently that the SCUBA
bolometer array on the James Clerk Maxwell Telescope (JCMT) has
provided  sufficient sensitivity at submillimeter  (sub-mm)
wavelengths (450, $\rm 850 \mu$m) to directly confirm the presence of a
dominant cold dust component in  nearby spirals (Alton et al. 1998a;
Alton et al. 1998b; Frayer et al. 1999; Papadopoulos \& Seaquist 1999;
Dunne et al. 2000; Dunne \& Eales 2001).   

In more active luminous infrared galaxies (LIGs), with enhanced
star-formation or the presence of a central AGN, sub-mm observations
also suggest the presence of cold cirrus. Unlike  quiescent spirals
however, LIGs also show a prominent warm dust component (T=30--60\,K)
most likely heated  by the starburst (Klaas et al. 2001; Lisenfeld,
Issak \& Hills 2000). The enhanced activity in LIGs is believed to be
triggered by  interactions and mergers. It is therefore, interesting
to explore the  role of such violent events in heating the dust and
hence, modifying the relative amounts of the cold and the warm dust
component in galaxies.   Klaas et al. (2001) have searched for
systematic differences in  the far-infrared (far-IR) to sub-mm
luminosity ratio of interacting and non-interacting LIGs classified on
the basis of their optical morphology. Their analysis did not show any
trends suggesting either problems with their classification scheme or
similar dust temperature distributions in the two sub-samples. Studies
of  selected interacting galaxies (mostly LIGs) indeed, suggest that
cold dust is present in these systems even in the regions experiencing 
starburst activity where the stellar radiation field heating the  dust
is most intense (Haas et al. 2000). This may be due to effective 
shielding of the cold dust component by warmer dust. 

In this paper we further explore the issue of dust heating during galaxy
encounters by exploiting the large SCUBA database as well as
sub-mm observations presented in the literature. Our interacting
galaxy sample is compiled from the Georgakakis et al. (2000) study and 
comprises well separated spirals, systems close to nuclear coalescence
and young merger remnants. Our sample is larger compared to previous
studies and also spans a wider range of interactions stages.
%including early interacting systems.   

Section \ref{sample} discusses our sample selection while  Section
\ref{observations}  presents the reduction of the data used in our
study.  In Section \ref{results} we present  our results which are
discussed in Section \ref{discussion}. Finally section
\ref{conclusions} summarises  our conclusions. Throughout this paper
we adopt $\rm  H_{o}=75\,km\,s^{-1}\,Mpc^{-1}$.    

%%%%%%%%%%%%%%%%%%%%%%%%%%%%%%%%%%%%%%%%%%%%%%%%%%
\section{The sample}\label{sample}
%%%%%%%%%%%%%%%%%%%%%%%%%%%%%%%%%%%%%%%%%%%%%%%%%%

In this paper we use the interacting galaxy sample presented by Georgakakis 
et al. (2000) comprising disk galaxy mergers of similar mass. This is culled 
from a number of studies (Keel \& Wu 1995; Gao \& Solomo 1999; Gao et
al. 1998;  Surace et al. 1993) in an attempt to minimize any biases
introduced by different  selection criteria (e.g. far-IR,
morphological selection) used to compile  interacting galaxy
samples. We note however, that most of the galaxies in that study are
far-IR luminous and are also biased against mergers occurring along
our line of sight. Therefore, although the Georgakakis  et al. (2000)
sample is  by no means statistically complete, it could  be regarded
as representative of  interacting systems and merger remnants spanning
a wide range of properties.  The sample employed in this study is
presented in Table 1 and comprises only  those systems from the
Georgakakis et al. (2000) study that have sub-mm data  from either the
SCUBA archive or the literature.   

%%%%%%%%% FIG 1
\begin{figure*}
\epsfig{figure=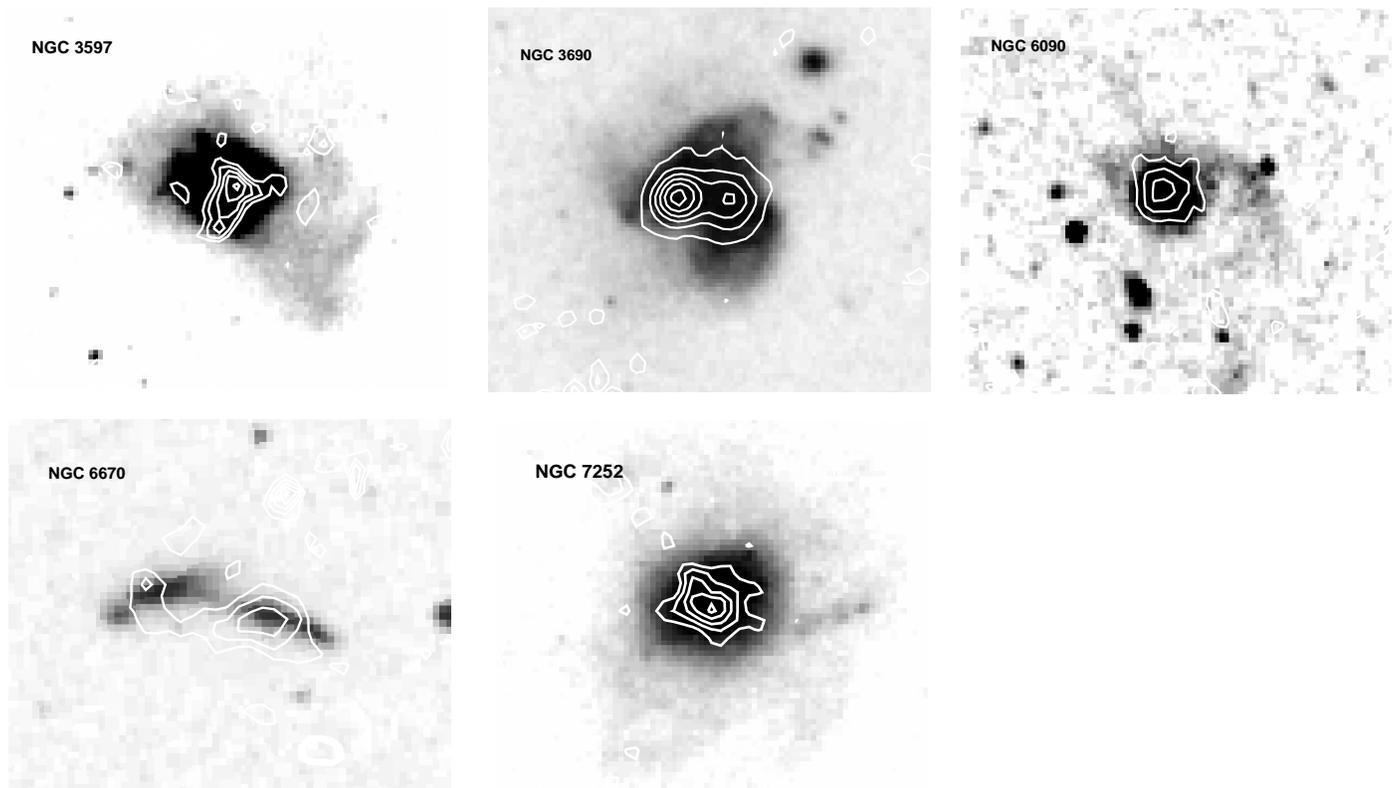, width=18.5truecm, angle=0}
%bbllx=0pt, bblly=50pt,bburx=580pt,bbury=720pt}
\caption[]{Sub-mm images of NGC 3597, NGC 3690, NGC 6090, NGC 6670 and NGC 7252.
Contour levels are 0.020, 0.033, 0.047, 0.060, 0.073 Jy/beam for NGC 3597;
0.030, 0.068, 0.107, 0.146, 0.184, 0.222 Jy/beam for NGC 3690;
0.020, 0.051, 0.082 Jy/beam for NGC 6090;
0.040, 0.063, 0.086 Jy/beam for NGC 6670 and
0.010, 0.017, 0.025, 0.032, 0.040 Jy/beam for NGC 7252.}
\label{f1}
\end{figure*}

Following Georgakakis et al. (2000) we use the galaxy `age' parameter 
providing an estimate of the evolutionary stage of each system  relative 
to the time of the merging of the two nuclei. Such a chronological sequence 
allow us to explore the evolution of the galaxy properties at different times
during the interaction. Negative `ages' are for pre-mergers  while
positive `ages' correspond to merger remnants. For pre-mergers the 
`age' is estimated by dividing the projected separation of the two nuclei,
$\delta r$, by an (arbitrary) orbital decay velocity
$v=30\,\mathrm{km\,s^{-1}}$.  It is clear that the `age' parameter for 
pre-mergers is affected by projection effects or different interaction
geometries. However, to the first approximation, it provides an estimate of
the stage of the merging and allows plotting of pre- and post-mergers on
the same scale. For post-mergers we adopt the evolutionary sequence defined by  
Keel \& Wu (1995) using dynamical and morphological criteria. In particular, 
the `age' parameter for these systems is calculated by multiplying the 
dynamical stage number, defined by  Keel \& Wu (1995), by the factor
$4\times10^{8}$\,yr. This conversion factor is found to be 
appropriate for  the 3 merger remnants in the Keel \& Wu sample with
available spectroscopic estimates (i.e. NGC 2865, NGC 3921, NGC 7252;
Forbes, Ponman \& Brown 1998). It should be stressed that the `age'
parameter for both pre- and post-mergers does not represent an absolute
galaxy age but is an indicator of the evolutionary stage of the
interaction. 

%%%%%%%%%%%%%%%%%%%%% TABLE 1 %%%%%%%%%%%%%%%%%%%%
\begin{table*}
%\begin{minipage}{8.7truecm}
%\begin{center}
\caption{Observational information of the galaxies.} 
\begin{tabular}{l cc cc c l l l} 

\hline

Name & age & $D$ &  $f_{60}$ & $f_{100}$ & $f_{850}$ & $\lambda$ & $f_\lambda$ & Refs.\\
     & ($\rm \times 10^8\, yr$) & (Mpc) & (Jy) & (Jy) & (Jy) & ($\rm \mu m$) & (Jy) & \\
\hline
         ARP\,220 &  $+0.0$ &  83.6 &103.33  & 113.95 & $0.832\pm0.086$ & 
 350, 450, 800 & 11.7, 3.0, 0.83  & f, c, c\\
	         &          &       &        &        &                 & 
       1100, 1250, 1300  & 0.35, 0.23, 0.02 & f, d, a\\
         ARP\,240 & $-11.2$ &  90.6 & 10.68  &  18.69 & $0.283\pm0.039$ &
         &  &  \\
         ARP\,271 &  $-8.0$ &  34.9 &  9.93  &  24.81 & $0.548\pm0.086$ &
         &  &  \\
        ARP\,302 &  $-8.8$ & 134.7 &  6.68  &  14.54 & $0.215\pm0.031$ &
       450, 1100& 0.37, 0.04 & e, e \\
          ARP\,90 &  $-1.9$ &  33.2 &  9.14  &  13.69 & $0.119\pm0.022$ &
         &  & \\ 
          ARP\,91 &  $-1.9$ &  28.8 & 11.55  &  19.50 & $0.306\pm0.031$ &
         &  & \\
         IC\,1623 &  $-2.0$ &  80.2 & 22.19  &  30.32 & $0.273\pm0.038$ &
   450  & 2.43 & h\\
          IC\,883 &  $-0.1$ &  93.3 & 13.69  &  24.90 & $0.104\pm0.016$ &
  350, 750 & 1.95, 0.20 & j, j \\ 
 IRAS\,01418+1651 &  $-1.4$ & 109.7 & 11.86  &  13.75 & $0.076\pm0.015$ &
         &  & \\
 IRAS\,03359+1523 &  $-2.2$ & 141.5 &  5.77  &   6.53 & $0.044\pm0.009$ &
         &  & \\
         MRK\,848 &  $-1.6$ & 160.7 &  9.15  &  10.04 & $0.060\pm0.014$ &
   450   &  0.42 & j\\
         MRK\,273 &  $-0.2$ & 151.0 & 27.45  &  22.44 & $0.104\pm0.010$ &
  350, 450, 800  & 1.0, 0.71, 0.08  & f, f, f\\
                  &         &       &        &        &                 & 
        1100, 1300 & 0.05, 0.02 & f, b\\
        NGC\,1614 &  $-0.7$ &  63.7 & 33.12  &  36.19 & $0.219\pm0.032$ &
   350, 450 & 1.87, 0.98 & g, j\\
        NGC\,2623 &  $+1.6$ &  73.8 & 25.72  &  27.36 & $0.091\pm0.014$ &
   350, 750 & 2.25, 0.17 & g, j\\
        NGC\,3597 &  $-0.3$ &  46.5 & 12.71  &  15.96 & $0.061\pm0.006$ &
          & & \\
        NGC\,3690 &  $-1.4$ &  40.6 &121.64  & 122.45 & $0.486\pm0.049$ &
        350, 450, 1250 & 7.50, 1.50, 0.10 & a, l, d\\
   NGC\,4038/4039 &  $-2.5$ &  21.9 & 48.68  &  82.04 & $0.760\pm0.076$ &
        450 & 4.08 & i\\
        NGC\,4194 &  $+0.8$ &  33.4 & 25.66  &  26.21 & $0.113\pm0.220$ &
        350, 1300 & 0.97, 0.01 & g, b\\
        NGC\,4922 &  $-3.7$ &  98.1 &  6.20  &   7.30 & $0.053\pm0.012$ &
        & & \\
         NGC\,520 &  $-1.9$ &  30.4 & 31.55  &  46.56 & $0.325\pm0.050$ &
       350 & 4.89 & g\\
        NGC\,5256 &  $-1.6$ & 111.4 &  9.72  &  10.35 & $0.082\pm0.017$ &
       1300 & 0.01 & b\\
        NGC\,6052 &  $-0.5$ &  62.9 &  6.46  &  10.18 & $0.095\pm0.015$ &
       450 & 0.72 & j \\
        NGC\,6090 &  $-1.3$ & 117.0 &  6.25  &   9.34 & $0.103\pm0.010$ &
            150, 200& 8.67, 4.48 & m, m\\
        NGC\,6240 &  $-0.3$ &  97.9 & 32.22  &  22.71 & $0.155\pm0.045$ &
        450 & 1.07 & k \\
       NGC\,6670 &  $-4.9$ &  63.7 &  8.24  &  15.18 & $0.137\pm0.014$ &
          180, 200 & 14.5, 13.5& n, n\\
        NGC\,7252 &  $+2.4$ &  62.5 &  4.28  &   7.73 & $0.045\pm0.005$ &
         1250 &0.025&o \\
        NGC\,7592 &  $-1.9$ &  97.7 &  9.30  &  10.50 & $0.108\pm0.019$ &
         150, 180, 200& 6.4, 5.8, 4.6 & p, p, p\\
        UGC\,2369 &  $-4.3$ & 124.7 &  8.14  &  11.10 & $0.072\pm0.013$ &
        450 & 0.52 & j\\
        UGC\,4881 &  $-3.5$ & 157.1 &  6.53  &  10.21 & $0.065\pm0.013$ &
           & & \\
\hline
\multicolumn{9}{l}{
a: Chini et al. 1986;
b: Kruegel et al. 1988;
c: Eales, Wynn-Williams \& Duncan 1989;
d: Carico et al. 1992;
}\\
\multicolumn{9}{l}{
e: Clements, Andreani \& Chase 1993;
f: Rigopoulou, Lawrence \& Rowan-Robinson 1996;
g: Benford et al. 1999;
h: Frayer et al. 1999;
}\\
\multicolumn{9}{l}{
i: Haas et al. 2000;
j: Dunne \& Eales 2001;
k: Klaas et al. 2001;
l: this work.;
m: Calzetti et al. 2000;
n: Spinoglio, Andreani \& Malkan 2002;
}\\
\multicolumn{9}{l}{
o: Andreani \& Franceschini 1996;
p: Siebenmorgen, Krugel \& Chini  1999}\\
\end{tabular}
%\end{center}
%\end{minipage}
\end{table*}
%%%%%%%%%%%%%%%%%%%%%%%%%%%%%%%%%%%%%%%%%%%%%%%%%%%%%%%%%%%%%%%%%%%%%%

%%%%%%%%%%%%%%%%%%%%%%%%%%%%%%%%%%%%%%%%%%%%
\section{Observations}\label{observations}
%%%%%%%%%%%%%%%%%%%%%%%%%%%%%%%%%%%%%%%%%%%%
 
In addition to the sub-mm data available in the literature we have 
analysed unpublished SCUBA archival data for five merging systems in the
present sample: NGC\,3597,  NGC\,3690, NGC\,6090, NGC\,6670
and NGC\,7252. The observations of these galaxies were made with the
SCUBA sub-mm bolometer array (Holland et al. 1999)  which is mounted
on the JCMT and provides simultaneous imaging at 450 and 850\,$\rm \mu m$
for a region of sky of 2.3\,arcmin in diameter. SCUBA is an array of
37 bolometers (HPBW=14.7\,arcsec) at 850\,$\rm \mu m$ and 91
(HPBW=7.5\,arcmin) at 450\,$\rm \mu m$. In order to provide fully sampled
images, the secondary mirror moves in a 64-step jiggle pattern, with
the integration time lasting 1\,s at each position. After the 16 steps
of jiggle pattern, the telescope nods in order to allow for slowly
varying sky gradients. For NGC\,3597  (observed in 5-Apr-2000),
NGC\,3690 (observed in 24-Nov-98), NGC\,6670 (observed in 5, 6-Nov-97)
and NGC\,7252 (observed in 5-Jul-97)  the chop throw was 120, 80, 80
and 120\,arcsec respectively. For NGC\,6090 a chop throw of 120, 80
and 60\,arcsec was used during the observing runs of 12-Mar-1998, 
5-Dec-1997 and 4-Jan-2000 respectively.  All observations were made
with grade-1 weather ($\tau_{225} < 0.05$) except for 5-Apr-2000 and
4-Jan-2000 where weather conditions were of grade-3  ($\tau_{225} \sim
0.1$). 

The data were reduced using the SURF software package (Jenness et
al. 1997). We made corrections for the atmospheric absorption using
the opacities derived from skydip measurements taken regularly every
night. Noisy bolometers were flagged and large spikes were
removed. During each night, several calibration sources were mapped in
order to flux calibrate our images. The SCUBA beams have moderate
error lobes  and since all our targets are extended sources, we
corrected the calibration in order to account of emission outside the
central beam. 
We estimated the flux loss due to the secondary lobes by
evaluating the average peak-to-aperture flux ratios measured for the
calibration sources for each observing night. All five new galaxy
systems presented in this work were detected at 850\,$\rm \mu m$ while
only NGC\,3690 showed significant signal in the 450\,$\rm \mu m$ band. For
all our observations we estimated a calibration uncertainty (based
on differences among the nightly calibration values) which are
reported in Table 1.
The $850\,\rm \mu m$ emission of these galaxies is presented in
Figure \ref{f1}  with contours overlaid on top of the optical DSS
images. Unfortunately, the relatively large beam size of SCUBA ($\sim
4$\,kpc at a distance of 50 Mpc) does not allow for a morphological
study of these systems. We note however, that some of these systems
(mostly early pre-mergers) have $850\,\rm \mu m$ emission that shows
two peaks, while others (the more evolved systems) show a single 
peak. 

%%%%%%%%%%%%%%%%%%%%%%%%%%%%%%%%%%%%%%%%%%%%%%%%%%
\section{Results}\label{results}
%%%%%%%%%%%%%%%%%%%%%%%%%%%%%%%%%%%%%%%%%%%%%%%%%%

The dust properties of our sample are quantified using the
100--to--$\rm  850\, \mu m$ flux density ratio,
$f_{100}/f_{850}$. This provides an estimate of the relative mass
fraction of the warm and the cold dust component. Indeed, the sub-mm
wavelengths (e.g. 450 and $\rm 850\, \mu m$) are sensitive to cold
dust ($\rm T\approx15-25\,K$; Alton et al. 1998a; Alton et al. 1998b;  
Bianchi et al. 1998; Dunne et al. 2000; Dunne \& Eales 2001) while the 
$\rm 100\, \mu m$ flux density primarily probes the warmer dust ($\rm
T\approx35\,K$) heated by the star-formation activity (Devereux \&
Young 1990; Devereux et al. 1994;  Klaas et al. 2001; Misiriotis et
al. 2004).  Also the presence of an AGN in some of our systems 
does not affect our results. This is because  at the wavelengths used
in this study the contamination of the observed fluxes due to  AGN
emission is expected to be small (e.g. Della Ceca et al. 2002; Zezas,
Ward \& Murray 2003). This is also discussed below.

We further explore the connection between $f_{100}/f_{850}$ and the
dust content by estimating the warm ($\rm M_W$) and the cold
($\rm M_C$)  dust mass of our sample galaxies and then plotting
$f_{100}/f_{850}$ against the ratio  $\rm M_C /
M_W$. The $\rm M_W$, $\rm M_C$ are calculated by
fitting two modified Planck functions,  each with different
temperature, to the observed far-IR to sub-mm spectral energy 
distribution of our sources. We follow the method described by Dunne \&
Eales (2001) keeping the dust emissivity index fixed to $\beta=2$. 
In this exercise we combine the IRAS and the SCUBA data presented in
Table 1 with the flux density measurements at other far-IR and sub-mm
wavelengths from  the literature. Only systems with data
in at least four wavelengths, of which at least two should lie in the
sub-mm regime,  are fitted.  

The emission at a particular wavelength is modeled by a two-temperature
graybody described by

\begin{equation}
F_{\lambda} = \frac{N_W}{\lambda^{\beta}} B(\lambda, T_W) + 
\frac{N_C}{\lambda^{\beta}} B(\lambda, T_C),
\end{equation}  

\noindent
where $T_W$ and $T_C$ are the temperatures of the warm and the cold dust
component respectively and $\beta$ is the dust emissivity index, fixed
to $\beta=2$ for consistency with the study of Dunne \& Eales 2001. 
$N_W$ and $N_C$ are normalization constants that determine the
mass of each of the warm and cold components respectively. As already
argued in Alton et al. (1998a), although the dust masses that are 
calculated with this technique are not very accurate (due to the
adopted  dust properties), the cold--to--warm dust mass ratio 
$\rm M_C / M_W$ (which is equivalent to $\rm N_C / N_W$)
provides a more secure estimate of the relative content of warm and
cold dust within the galaxies.
We also estimate the statistical error of the $\rm M_C / M_W$
ratio due to the uncertainty in the measured fluxes using the
bootstrap method described by Dune \& Eales (2001). For each galaxy in 
our sample mock SEDs were constructed by randomising the fluxes at
each wavelength, assuming that they follow a Gaussian distribution
with mean the measured flux and standard deviation the uncertainty
estimated from the observations. A total of 100 mock SEDs were
produced for individual galaxies in our sample and each of them was
then fitted with two modified black bodies in the same way as with the
real data. The standard deviation of $\rm M_C / M_W$ ratio from the
100 trials provides an estimate of the uncertainty in this parameter.   
We note however, that this uncertainty does include systematic biases
that  may affect  the $\rm M_C / M_W$ ratio. For example,  the
sparse wavelength coverage, uncertainties in the opacity coefficients 
$\kappa_d$ and the emissivity index $\beta$ as well as degeneracies
in the fitting parameters make the decomposition of the observed SED
into two blackbodies not unique (Klaas et al. 2001).

\begin{figure}
\epsfig{figure=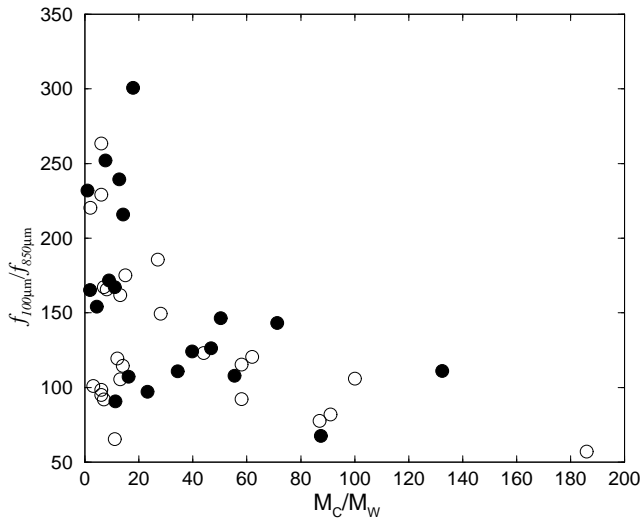, width=8.5truecm, angle=0}
\caption{$f_{100}/f_{850}$ flux ratio against the cold--to--warm
dust content. The filled circles are the interacting systems in the present
sample while the open circles are the galaxies presented in Dunne \& Eales
(2001).}
\label{fig_mwmc}
\end{figure}

The results of the fits are presented in Table 2. Figure
\ref{fig_mwmc} plots $f_{100}/f_{850}$  against the ratio  $\rm M_C /
M_W$ for the merging galaxies presented in this study (solid circles)
and the galaxies presented in Dunne \& Eales 2001 (open circles).  
Despite the large scatter there is evidence for a correlation in the
sense that sources with a lower cold--to--warm dust mass fraction (low 
$\rm M_C / M_W$ values) also have enhanced $f_{100}/f_{850}$ flux
density ratio. The Spearman's rank correlation  test rejects the null
hypothesis that the $f_{100}/f_{850}$ and the $\rm M_C  /  M_W$ are
uncorrelated with a probability 0.015 corresponding to $\approx
2.5\sigma$. Similar conclusions are  obtained if we use the larger
sample of Dunne \& Eales  (2001) that comprise both interacting
systems and isolated galaxies.  

The large scatter in Figure \ref{fig_mwmc} may be intrinsic suggesting
that parameters other than $\rm M_C / M_W$ may affect the $f_{100} /
f_{850}$. However, the systematics discussed above may also bias the $\rm
M_C / M_W$ ratio and contribute to the observed scatter. We avoid the
above systematics by using observables to describe the dust properties
(e.g. $f_{100} / f_{850}$) rather than parameters derived from
spectral fittings.  

%%%%%%%%%%%%%%%%%%%%%%%%%%%%%%%%%TABLE 2%%%%%%%%%%%%%%%%%%%%%%%%%%%%%%%%%
\begin{table}
\caption{Dust temperatures and cold--to--warm dust mass ratio
derived by fitting a two-temperature graybody curve with $\beta = 2$.
The uncertainty of the cold--to--warm dust mass ratio is given in
tha last coloumn.}
\begin{tabular}{l c c c c}
\hline
Name & T$_W$ (K) &  T$_C$ (K) & M$_C$/M$_W$ & $\sigma_{(\rm M_C / M_W)}$ \\
\hline
ARP\,220    & 49.3 & 22.8 & 39.7& 10.4\\
ARP\,302    & 40.9 & 20.1 & 87.4&13.6\\
IC\,1623    & 55.6  & 23.1 & 132.4&8.9 \\
IC\,883     & 43.2  & 17.0 & 12.7&2.5\\
MRK\,848    & 35.5  & 11.7 & 11.1&2.5 \\
MRK\,273    & 50.4  & 27.4 & 14.1&5.2 \\
NGC\,1614   & 38.7  & 23.1 & 1.8 &0.9\\
NGC\,2623   & 48.4  & 26.2 & 17.8&1.8\\
NGC\,3690   & 36.8  & 36.7  & 7.6&0.4\\
NGC\,4038/4039   & 21.4  & 28.6 & 55.4& 9.7\\
NGC\,4194   & 36.5  & 36.5 & 0.87&1.1 \\
NGC\,520    & 52.2  & 25.1 & 71.2&3.2 \\
NGC\,5256   & 48.4  & 23.1 & 46.7&18.7 \\
NGC\,6052   & 34.7  & 19.3 & 16.1&2.3 \\
NGC\,6090   & 32.3  & 14.7 & 11.2&2.0 \\
NGC\,6240   & 51.8  & 22.2 & 50.3&14.7 \\
NGC\,6670   & 30.9  & 14.2 & 34.4&6.0 \\
NGC\,7252   & 34.9 & 22.3 & 8.9&1.2 \\
NGC\,7592   & 33.6  & 14.2 & 23.1&2.9 \\
UGC\,2369   & 34.3  & 18.9 & 4.4&2.0 \\
\hline
\end{tabular}
\end{table}

\begin{figure}
\epsfig{figure=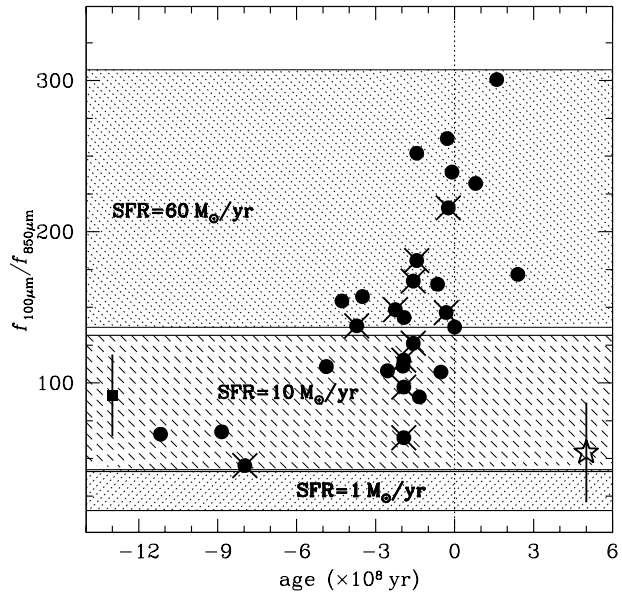, width=8.5truecm, angle=0}
\caption{$f_{100}/f_{850}$ flux ratio against the age parameter
for our sample of interacting galaxies. The filled circles are the
interacting galaxy sample in this study. A cross on top of symbol
signifies an AGN in the catalogue of Veron-Cetty \& Veron (2003). The
square corresponds to the mean $f_{100}/f_{850}$ of isolated spirals
from the sample of Misiriotis et al. (2004) arbitrarily plotted at an
age parameter of --13. The star is the mean $f_{100}/f_{850}$ of
ellipticals from the sample of Temi et al. (2003; see text for
details)   arbitrarily plotted at an age parameter of $+5$.} 
\label{fig_age}
\end{figure}

Figure \ref{fig_age} plots the ratio $f_{100}/f_{850}$ as a function
of the age parameter. As explained in Section 2, negative ages
correspond to pre-merger stages, zero is for the nuclear coalescence
while positive ages are for merger remnants.  In Figure
\ref{fig_age}  there is evidence for an increase in the
$f_{100}/f_{850}$ ratio as the interaction progresses toward the final
stages of the merger. We estimate a Spearman rank correlation
coefficient $r=0.64$ and a probability that there is no correlation
$P=0.07$ per cent. Therefore, the null hypothesis that the
$f_{100}/f_{850}$ and the age parameter are uncorrelated can be
rejected at the $\approx3.5\sigma$ confidence level.

We note that this statistically significant correlation is not
due to $f_{100}$ or  $f_{850}$ being individually correlated with the
age parameter. We normalise $f_{100}$ and $f_{850}$ with the $K$-band
flux ($f_K$) of our systems from the Two Micron All Sky Survey (2MASS;
Jarrett et al. 2000). Using the Spearman rank correlation test we find
that the probability that the age parameter is uncorrelated with
$f_{100}/f_K$ and $f_{850}/f_K$ of 4 ($\approx 2 \sigma$) and 28
($\approx 1 \sigma$) per cent respectively. Therefore, the null
hypothesis that there is no correlation between these parameters
cannot be rejected at a high confidence level.  Also, the trend
between $f_{100}/f_{850}$ and the age parameter is not affected by the
presence of an AGN in some of our sample galaxies. This is
demonstrated in Figure \ref{fig_age} where systems identified as AGN
in the catalogue of Veron-Cetty \& Veron (2003) are plotted with
crossed circles. The AGNs in that figure  follow a similar trend to
the remaining systems.

The mean $f_{100}/f_{850}$ ratio of isolated spirals from the sample
of  Misiriotis et al. (2004) is also plotted in Figure \ref{fig_age}
at an arbitrary  age of --13. Based on the discussion above this trend 
can be interpreted as an increase of the warm relative to the cold
dust mass from isolated spirals and early interacting systems to
advanced mergers. Such a transition of the cold dust to warmer
temperatures can be attributed to heating by the more intense UV
radiation field of the enhanced star-formation triggered by  the
merging (e.g. Casoli et al. 1991; Keel \& Wu 1995; Georgakakis et
al. 2000). However, the lack of a statistically significant correlation
between  $f_{100}/f_K$ ($\approx 2\sigma$; providing an estimate of
the SFR normalised to the total galaxy mass) and age suggests that
parameters other than the SFR also play an important role in
moderating the  $f_{100}/f_{850}$ ratio.

We further explore this  using the sophisticated model of
Misiriotis et al. (2004) for the absorption of the UV/optical
radiation by dust and its re-emission at far-IR and sub-mm
wavelengths. This model produces UV to sub-mm SEDs for a given input
system geometry, stellar population (e.g. young and old stars), dust
content and total galaxy mass. Our goal is not to derive physical
parameters for the individual interacting systems but to show that
this model, under simple assumptions, can reproduce the observed
$f_{100}/f_{850}$ range of our sample.  

The interacting galaxies studied here clearly have complex morphology 
that is difficult to quantify and to model. To the first approximation
we assume simple exponential disk-galaxy geometry. Although clearly
a more complex geometry is required for each one of our interacting
systems, this is an approximation that 
allows some insight into how different parameters (e.g. star-formation
rate, dust content) affect the $f_{100}/f_{850}$ ratio. Since we do
not attempt to model individual systems the exact geometry of the
interaction is not a crucial parameter for this study. The total mass
of the model galaxy is parametrised by its $K$-band luminosity. We
adopt $\rm M_K=-24.5$\,mag, the mean absolute magnitude of our sample 
galaxies. The dust mass of the fiducial galaxy is assumed to be in the
range $\rm \approx 10^7 - 10^8\,M_{\odot}$, similar to the dust masses  
estimated using the simple black-body fits described above. The
stellar content is directly associated with the density of the UV
radiation responsible for the heating of the dust and is parametrised
by the star-formation rate. Here we use three different values for the 
star-formation rate  SFR=1, 10 and $60\, M_{\odot}$. For a given
SFR and dust mass (in the range  $\rm \approx 10^7 - 10^8\,M_{\odot}$) we
derive the $f_{100}/f_{850}$ ratio predicted by the Misiriotis et
al. (2004) model. The shaded regions in Figure \ref{fig_age}
represent the $f_{100}/f_{850}$ for different SFRs. The lower and
upper boundaries of these regions correspond to dust masses of $10^7$
and $10^8\, M_{\odot}$  respectively. Clearly, both the SFR and the
total dust mass (relative to the stellar mass) are important for the
$f_{100}/f_{850}$ ratio.

As expected the model predicts that higher star-formation rates heat
the dust resulting in  higher $f_{100}/f_{850}$ ratios. Indeed, the
model  of Misiriotis et al. (2004) predicts an increase in the
mean (weighted by mass) dust temperature of the cold component
for different SFRs: we find that  for SFR=1\,$M_{\odot}$/yr the dust
temperature is between 16 and 18\,K, for 10\,$M_{\odot}$/yr  it is
between 19 and 22\,K while for SFR=60\,$M_{\odot}$/yr it is between 24
and  28\,K.

Also, keeping the SFR fixed, more dusty systems have lower
$f_{100}/f_{850}$ which can be attributed to higher fraction of
UV/optical photons absorbed by the dust, heating it to temperatures
$T \la 30$\,K and then being re-emitted primarily at sub-mm
wavelengths (e.g. $\rm 850\mu m$). 
For example in the case of  the Misiriotis et al. (2004) study, the
fraction of UV/optical photons absorbed by the diffuse dust component
in individual systems in their sample (as predicted by their best fit
model) is anti-correlated with the observed $f_{100}/f_{850}$  at the
$2.5\sigma$ confidence level.  This may also partly explain the
scatter seen in Figure \ref{fig_age} since the relative fraction of
the warm--to--cold dust for a given SFR varies substantially with dust
content. Other factors such as projection effects, differences in the
details of individual interactions (e.g. geometry, initial conditions,
bulge--to--disk ratio) may also contribute to the scatter in Figure
\ref{fig_age}. We also note that the diffuse UV/optical radiation
field does not have a large impact on the $\rm 100\mu m$ emission,
which is primarily dominated by starburst activity taking place in
localised galaxy regions.

%%%%%%%%% FIG 2
%%%%%%%%% FIG 3

At post-merger stages the evolution of $f_{100}/f_{850}$ with age
after the merging event remains poorly constrained since only 3
systems have SCUBA measurements available. This is also a problem for
evolved ellipticals, the alleged product of disc galaxy mergers,
with dust properties (e.g. temperature, content) that remain largely
unknown. This is mainly because these systems are believed to have
very little if any dust and hence are difficult to detect at far-IR
and sub-mm wavelengths. Previous observations of ellipticals at sub-mm
wavelengths targeted far-IR luminous and optically peculiar systems
to maximize the probability of detection (e.g. Fich \& Hodge 1993;
Wiklind \& Henkel 1995; Leeuw et al. 2004). As a result the
$f_{100}/f_{850}$ ratios of these systems may not be representative of
ellipticals. We attempt to constrain the $f_{100}/f_{850}$ ratio of
evolved ellipticals using the sample of Temi et al. (2003). These
authors use ISO observations to constrain the far-IR (60--200\,$\rm \mu
m$) SEDs of early type galaxies (not selected on the basis of their
far-IR luminosity) and to estimate their warm and cold dust content
using methods similar to those outlined here (e.g. modified black-body
fittings). Although their results are  somewhat uncertain due to the
lack of sub-mm observations which are critical as a probe of the cold
dust, this study provides the only constraints on the dust properties
of early type systems using data extending beyond $100\rm \mu m$. Using
the best fit model of Temi et al. (2003) we extrapolate their observed
SEDs to $\rm 850\, \mu m$ and estimate the mean $f_{100}/f_{850}$ flux
density ratio for the elliptical galaxies in their sample. This mean
value ($54 \pm 33$) is plotted in Figure \ref{fig_age} (the star
arbitrary plotted at an age of $+5$)  and should be considered with
caution since the $f_{850}$ flux density is not directly observed.
Nevertheless, taken at face value it suggests that the ratio
$f_{100}/f_{850}$ in evolved ellipticals  is about 1\,dex lower
compared to young merger remnants. This can be interpreted as dust
cooling after the merger event under the hypothesis that
elliptical galaxies are formed via disc galaxy mergers. We caution the
reader however, that other mechanisms such as spattering of dust
grains by X-ray radiation as well as the dispersal of dust in star
formation may also affect the  $f_{100}/f_{850}$ ratio. Observations
at sub-mm  wavelengths of elliptical  galaxies and post-mergers are
essential to further explore different scenarios.

\section{Discussion}\label{discussion}

In this paper we use an interacting galaxy sample ordered into a
chronological merging sequence to explore dust properties during
gravitational galaxy encounters. We find evidence for an increase in
the flux density ratio $f_{100}/f_{850}$ from isolated spirals and
early interacting systems to intermediate and late stage mergers close
to nuclear coalescence. We interpreted this trend as an increase of
the warm relative to the cold dust mass. Both the star-formation
triggered by the merging and the dust content of individual systems
are important parameters for modifying the   $f_{100}/f_{850}$ ratio
and the observed trend along the merging galaxy sequence.

Klaas et al. (2001) have also investigated the far-IR to sub-mm
properties of luminous infrared galaxies and have searched for
systematic differences between isolated and interacting
systems classified on the basis of their optical morphology. Unlike
our results they find no differences in the relative amounts of warm
and cold dust between the two samples. However, these authors do not
attempt to order their interacting systems on the basis of the angular
separation between the two nuclei (ie. the age parameter). Also, their sample spans a small
range in bolometric luminosity and therefore well separated early
interacting systems may be underrepresented.  

Dunne \& Eales (2001) have performed the most extensive study of the
sub-mm dust properties of nearby galaxies todate using the SCUBA
bolometer array at the JCMT. They find that in quiescent spirals most
of the warm dust ($\rm T > 30\,K$) is concentrated in nuclear regions
with the cold dust component ($\rm T = 15 - 20\,K$) dominating the
outer disc.  The situation is strikingly different in starbursts where
they find that the  warm dust region is more extended
and therefore dominates the SED. This is attributed to
the more intense interstellar radiation field in actively star-forming
galaxies. Although Dunne \& Eales (2001) do not discuss possible
differences in the sub-mm properties between interacting and isolated
galaxies in their sample, most of our interacting/merging systems,
particularly the more advanced ones, are in a starburst
phase. Therefore, on the basis of the Dunne \& Eales (2001) findings,
for these galaxies we would expect a significant warm dust component
compared to isolated (on average more  quiescent) spirals. 

We note however, that the evidence above does not suggest the absence
of cold dust in interacting starbursts. Indeed, detailed study of
selected systems (e.g. ARP\,220, NGC\,4038/4039) shows significant
amounts of cold dust that may dominate the total dust content (Klaas
et al. 1997; Haas et al. 2000: Klaas et al. 2001; Dumke, Krause \&
Wielebinski 2003). Dunne \& Eales (2001) argue that the most active
starburst activity is usually taking place in well-shielded  dust
enshrouded regions allowing the presence of very cold dust at the same
regions.    

Moreover, using our interacting galaxy sample we attempt to explore
possible connections between the dust properties of isolated spirals,
mergers and evolved ellipticals. This is particularly important in the
light of the merger hypothesis postulating that ellipticals are the
product of disc galaxy mergers. However, our sample comprises only a
few post merger remnants while the cold and warm dust properties
of evolved ellipticals remain poorly constrained. Nevertheless, the
tentative result of our study is that young merger remnants have
$f_{100}/f_{850}$ ratios about 1\,dex higher than evolved ellipticals.
This may suggest that the dust cools with time after nuclear
coalescence, although other scenarios can also explain the lower
$f_{100}/f_{850}$ ratios in these systems. Future observations using
sensitive instruments such as ALMA and the {\it Spitzer}  satellite
will allow us to probe the largely unexplored far-IR to sub-mm regime
of early type galaxies. These data have the potential to accurately
characterize, for the  first time,  the dust properties of ellipticals
and merger remnants and to search for possible evolutionary effects
between these systems.  

%%%%%%%%%%%%%%%%%%%%%%%%%%%%%%%%%%%%%%%%%%%%%%%%%%
\section{Conclusions}\label{conclusions}
%%%%%%%%%%%%%%%%%%%%%%%%%%%%%%%%%%%%%%%%%%%%%%%%%%

In this paper we explore the cold and warm dust properties of galaxies
during gravitational encounters using a merging galaxy sample ordered
into a time sequence and comprising well separated spirals, systems
close to nuclear coalescence and young merger remnants. We use
the 100  to $\rm 850\mu m$ flux density ratio,  $f_{100}/f_{850}$, as a
proxy to  the mass fraction of the warm and the cold dust in these
systems. We find evidence for an increase in $f_{100}/f_{850}$ along
the merging sequence from isolated spirals and early interacting
systems to advanced mergers.  We interpret this trend as an increase
of the warm relative to the cold dust mass and argue that the
star-formation and the dust content of individual systems are
important for modifying the  $f_{100}/f_{850}$ ratio.

%%%%%%%%%%%%%%%%%%%%%%%%%%%%%%%%%%%%%%%%%%%%%%%%%%%%%%%

\section{Acknowledgments}
We thank the referee, L.  Dunne, for useful comments and suggestions
that significantly improved this paper. AG acknowledges funding by the
European Union and the Greek Ministry of Development in the framework
of the Programme  `Competitiveness-Promotion of Excellence in
Technological Development and Research-Action 3.3.1', Project `X-ray
Astrophysics with ESA' s mission XMM', MIS-64564.

\end{document}